\documentclass[12pt]{article}
\usepackage{epsfig,amssymb,amsmath}
\title{Bounds on the density of sources \\
of ultra-high energy cosmic rays \\
from the Pierre Auger Observatory 
}
\begin{document}
\maketitle
\par\noindent
{\bf The Pierre Auger Collaboration} \\
P.~Abreu$^{61}$, 
M.~Aglietta$^{49}$, 
M.~Ahlers$^{91}$, 
E.J.~Ahn$^{79}$, 
I.F.M.~Albuquerque$^{15}$, 
I.~Allekotte$^{1}$, 
J.~Allen$^{83}$, 
P.~Allison$^{85}$, 
A.~Almela$^{11,\: 7}$, 
J.~Alvarez Castillo$^{54}$
% ...

\par\noindent
$^{1}$ Centro At\'{o}mico Bariloche and Instituto Balseiro (CNEA-UNCuyo-CONICET), San 
Carlos de Bariloche, 
Argentina \\
% ...
$^{7}$ Instituto de Tecnolog\'{\i}as en Detecci\'{o}n y Astropart\'{\i}culas (CNEA, CONICET, UNSAM), 
Buenos Aires, 
Argentina \\
% ...
$^{11}$ Universidad Tecnol\'{o}gica Nacional - Facultad Regional Buenos Aires, Buenos Aires,
Argentina \\
% ...
$^{15}$ Universidade de S\~{a}o Paulo, Instituto de F\'{\i}sica, S\~{a}o Paulo, SP, 
Brazil \\
% ...
$^{49}$ Istituto di Fisica dello Spazio Interplanetario (INAF), Universit\`{a} di Torino and 
Sezione INFN, Torino, 
Italy \\
% ...
$^{54}$ Universidad Nacional Autonoma de Mexico, Mexico, D.F., 
Mexico \\
% ...
$^{61}$ LIP and Instituto Superior T\'{e}cnico, Technical University of Lisbon, 
Portugal \\
% ...
$^{79}$ Fermilab, Batavia, IL, 
USA \\
% ...
$^{83}$ New York University, New York, NY, 
USA \\
% ...
$^{85}$ Ohio State University, Columbus, OH, 
USA \\
% ...
$^{91}$ University of Wisconsin, Madison, WI, 
USA \\
% ...
\par\noindent
(\ddag) Deceased \\
(a) Now at Konan University \\
(b) Also at the Universidad Autonoma de Chiapas on leave of absence from Cinvestav \\
(c) Now at University of Maryland \\
(d) Now at NYU Abu Dhabi \\

\begin{abstract}
We derive lower bounds on the density of sources of ultra-high energy cosmic rays from the lack of significant clustering in the arrival directions of the highest energy events detected at the Pierre Auger Observatory. 
The density of uniformly distributed sources of equal intrinsic intensity was found to be larger than $\sim (0.06 - 5) \times 10^{-4}$ Mpc$^{-3}$ at 95\% CL, depending on the magnitude of the magnetic deflections.
Similar bounds, in the range $(0.2 - 7) \times 10^{-4}$ Mpc$^{-3}$, were obtained for sources following the local matter distribution.

\end{abstract}

\section{Introduction}

Even many decades after the discovery of Ultra-High Energy Cosmic Rays (UHECRs),
their sources remain elusive.
The study of UHECR arrival directions is likely to provide significant 
insight into the still open question of their origin.
The trajectories of charged cosmic rays, protons and heavier nuclei, may be significantly bent by intervening galactic and extragalactic magnetic fields, thus losing correlation with their sources. 
However, UHECR arrival directions are the most likely to trace their sources, as the magnetic deflections are inversely proportional to the cosmic ray energy.

Also, the UHECR flux from distant sources is expected to be strongly attenuated
by the cosmic ray interactions with the cosmic microwave background, including photo-pion production for ultra-high energy protons (the so-called GZK effect) and photo-disintegration for heavier nuclei \cite{GZK}. 
Hence, cosmic rays of energy above $\sim 60$ EeV (1~EeV~$\equiv10^{18}$~eV) should mostly come from nearby sources, closer than about $200$ Mpc (see Sec. 4). The flux suppression measured at the highest energies
\cite{spechires,specauger} is consistent with an extragalactic origin of
UHECRs and with an energy attenuation due to the interaction of cosmic rays with photon backgrounds.

In this context, the observation of clustering in the arrival directions of UHECRs may shed light on their origin. For small magnetic deflections of the UHECR trajectories, the amount of clustering should reflect the density of local sources. In fact, the smaller the number of sources, the larger will be 
the UHECR flux coming from each of them, increasing the clustering signal which can be measured through the 
number of observed cosmic ray pairs separated by an angular distance smaller than the spread due to magnetic deflections. 
A statistical analysis of the clustering 
may help in identifying the astrophysical sources of UHECRs, since 
different populations of astrophysical objects have different characteristic
densities $\rho$, ranging from $\rho \sim 10^{-3} - 10^{-2}$ Mpc$^{-3}$ for normal galaxies  \cite{kochanek01} down to $ \sim 10^{-5} - 10^{-4}$ Mpc$^{-3}$ for 
Active Galactic Nuclei (AGN)  with X-ray luminosity $L_X > 10^{43}$ erg 
s$^{-1}$ \cite{agndens} and 
$\sim 10^{-7}$ Mpc$^{-3}$ for rich clusters of galaxies with mass larger than $10^{15}~ {\mathrm M}_\odot$ \cite{carlberg97}.

Both regular and turbulent magnetic fields play important roles in determining the strength of UHECR clustering. 
The regular component of the magnetic field, ${\bf B}$, induces a deflection $\delta$ in the arrival direction of an UHECR of charge $Z$ and energy $E$ reaching Earth from a distance $L$ along a trajectory $\bf s$ :
$$
\delta \simeq 3^\circ \frac{70\,\rm{EeV}}{E/Z}\left|\int_0^L
\frac{{\rm d}{\bf s}}{2\,{\rm kpc}}\times\frac{{\bf B}}{2\,\mu{\rm G}}\right| .
$$
A turbulent magnetic field with $rms$ amplitude $B_{rms}$ and coherence length $L_c$ will introduce an additional spread $\delta_{rms}$ in the arrival direction :
$$
\delta_{rms} \simeq 0.5^\circ \frac{70\,\rm{EeV}}{E/Z} \frac{B_{rms}}{3\,\mu {\rm G}} 
\sqrt{\frac{L}{2\,{\rm kpc}} \frac{L_c}{50\,{\rm pc}}}.
$$
These deflections are normalized to typical values of magnetic field amplitudes and distances in the Galaxy.
 The magnitude of the 
deflection induced by extragalactic magnetic fields has large uncertainties, with 
estimates ranging from $\lesssim 1^\circ$ \cite{dolag} to $\sim 20^\circ$ \cite{asm}
for 100 EeV protons.

Since a detailed knowledge of the galactic and extragalactic
magnetic fields is still missing, we will not attempt to model their effects in this paper.
Rather, our bounds on the density of sources will be given as a function of the angular scale and will only apply at angular scales larger 
than the spread due to magnetic deflections.

The UHECR composition at the highest energy is also relevant for the interpretation of the results presented in this paper. 
The Pierre Auger Observatory has measured a change of the  average shower depth, $\langle X_{max}\rangle$,  compatible with a transition from  
a light composition at 1 EeV to a heavier composition at  35 EeV, the highest energy for which results using this technique have been 
reported \cite{augerxmax,augercomp}. Complementary composition observables derived 
from the surface detector data including the asymmetry of the signals in the surface detector stations and the depth profile of muon production points give similar results at the highest energies, extending the results up to 45 EeV and 65 EeV respectively \cite{icrccomp}. 
HiRes \cite{hirescomp} and Telescope Array \cite{tacomp} have measured $\langle X_{max}\rangle$ with larger statistical uncertainties and they allow a wider range of compositions, including a pure proton one. The interpretation of measurements of shower depths of maximum in terms of composition relies on extrapolations of hadronic interactions to energies beyond the regime where they have been tested experimentally.
The UHECR composition at energies above 60 EeV is still not established. If heavy nuclei dominate at the highest energies, their magnetic deflections may be larger than $30^\circ$ (the maximum angular scale considered in our analysis), in which case our bounds will not apply.

Estimates of the density of sources in the range $10^{-6}
 - {\rm few} \times 10^{-3}$ Mpc$^{-3}$ have been obtained using data from 
previous experiments  
under various assumptions on the sources and their distribution 
\cite{dubovsky00,fodor00,yoshi03,blasi04,sigl04,kachel05}. 
More recent studies based on the arrival directions of 27 UHECRs ($E>56$~EeV) detected by the Pierre Auger Observatory before August 2007 have led to an estimate of  $\rho  \sim 10^{-4}$  Mpc$^{-3}$ \cite{cuoco09,takami}. A recent analysis of the autocorrelation function of UHECRs by the Telescope Array experiment shows no significant departure from isotropy \cite{ta2012}.

In this paper, we present an 
autocorrelation analysis of the arrival directions of the highest energy 
events detected by the Pierre Auger Observatory until 31 December 2011.  Bounds on the density of UHECR sources, assumed to be of equal intrinsic intensity, were derived for two plausible spatial distribution hypotheses, a uniform distribution and one following the local matter density as traced by galaxies in the 2MASS Redshift Survey (2MRS)
catalog \cite{2mass}. 
A preliminary analysis of an earlier dataset can be found 
in \cite{dedomenico}.

\section{The Pierre Auger Observatory and the data set}
\label{dataset}

The Pierre Auger Observatory is located in the 
Province of Mendoza, Argentina, at the Pampa Amarilla site (35.1$^\circ$-- 35.5$^\circ$ S, 
69.0$^\circ$-- 69.6$^\circ$ W and a mean altitude of 1400 m a.s.l.) \cite{augernim}. 
The Surface Detector (SD)
consists of 1660 water-Cherenkov stations arranged over an area of 3000 km$^2$ in a triangular grid of 1.5 km spacing \cite{augersup}. 
The array is overlooked by 27 Fluorescence Detector (FD) telescopes located on hills at
four sites on its periphery \cite{augerfl}. The FD provides a calorimetric measurement of the primary cosmic ray energy by reconstructing the shower development in the atmosphere.
Selection criteria for SD events include requiring the SD station with the largest signal to be surrounded by at least five active stations at the time of the event, and the reconstructed shower core to be inside a triangle of active stations.  The corresponding SD trigger efficiency is 100\% for $E> 3$ EeV and zenith angle $\theta < 60^\circ$. 
The cosmic ray arrival direction is obtained from the times of 
arrival of the shower front particles measured by the SD stations, with an angular resolution better 
than 0.9$^\circ$ for $E >$ 10 EeV \cite{angres2009}. 
The SD signal at 1000 m from the shower core, determined from a fit of the signals of the SD stations in the event, is used as an estimator of the cosmic ray primary energy and the FD is used to calibrate the SD estimator. The energy resolution for
$E > 10$ EeV is $12\%$, mainly coming from shower to shower fluctuations, and the systematic uncertainty on the absolute energy scale is $22\%$ \cite{enres2011}.
In this paper we will consider events with energy thresholds of 60, 70 and 80 EeV and with zenith angles smaller than $60^\circ$ which were recorded by the Surface Detector between 1 January 2004 and 31 December 2011. There are 84, 43 and 22 events with energy above 60, 70 and 80 EeV respectively.

\section{The two-point correlation function and the analysis method
}\label{methods}

A standard tool for the study of clustering in astronomical arrival directions  
 is the two-point angular correlation function, $n(\alpha)$, which gives 
the number of pairs separated by an angle smaller than $\alpha$:
\begin{equation}
n(\alpha) = \sum_{i=2}^N \sum_{j=1}^{i-1} \Theta (\alpha - \alpha_{ij}),
\end{equation}
where $\Theta$ is the step function, and $\alpha_{ij}$ is the angular distance 
between events $i$ and $j$ of $N$ cosmic rays above an energy threshold $E_{\mathrm{thr}}$. 

The number of pairs $n(\alpha)$ above an energy threshold has uncertainties due to both the energy and angular resolution of the experiment. Events with true energy close to $E_{\mathrm {thr}}$ may or may not be selected among the $N$ events with highest energy depending on their measured energy. The uncertainty on the measured angular distance $\alpha_{ij}$ may also affect the determination of the number of pairs at a given angular scale. The uncertainty on the energy has the largest effect on the number of pairs for the number of events and experimental uncertainties in the present data set. In order to estimate the effect of experimental uncertainties on $n(\alpha)$ we generated 1000 pseudo data sets by randomizing the energy and direction of each measured event according to the corresponding uncertainty. For each of these pseudo data sets, we calculated $n(\alpha)$ for the $N$ highest energy events (with $N= 83, 43$ and 22, the number of events measured above each of the energy thresholds considered).  From the distribution of these $n(\alpha)$ the uncertainty in the number of pairs in the data can be derived. As an example, we show in Figure 1 the case corresponding to $E_{\mathrm {thr}}= 70$ EeV. The mean number of pairs in the pseudo datasets is plotted along with error bars attached to each point. These error bars correspond to a 68\% CL derived from the pseudo data sets with a 16\% probability to be above (or below) the bar.
The shaded band in Figure 1 represents the 68\% CL range of the expected number of pairs for an isotropic distribution of $N$ arrival directions of cosmic rays, derived from Monte Carlo simulations which properly included the detector exposure.

\begin{center}
  \begin{figure}[t]
\epsfig{file=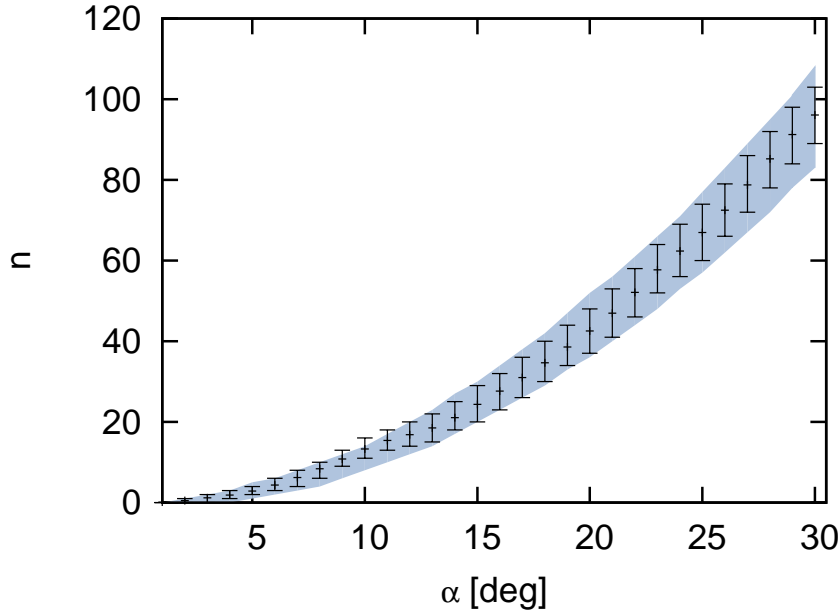,angle=270,width=12cm}
\caption{
 The two-point angular correlation function for an energy threshold of 70~EeV. The data points corresponds to $\bar n (\alpha)$ and the error bars to the
68\% CL  uncertainty in the energy and arrival directions. The shaded band represents the $68\%$ CL range of the expected number of pairs for an isotropic distribution of the arrival directions of cosmic rays as derived from Monte Carlo simulations.
\label{np70}}
  \end{figure}
\end{center}

The observed distribution of $n(\alpha)$ in Figure \ref{np70} is consistent with the expectation for an isotropic distribution of cosmic rays, and the lack of a significant clustering signal in the data can be used to establish a lower bound on the density of sources (the same happens also in the case of the 60 and 80 EeV energy thresholds). We use as a clustering estimator the mean number of pairs in the pseudo data sets described above, $\bar n (\alpha)$, and compare it with the results from Monte Carlo simulations of cosmic rays originating from different  distributions of sources with $N$ events above the energy threshold, taking into account the experimental uncertainties and the exposure of the Observatory. Details of the models for the distribution of sources  are given in Section \ref{subsec:density}.  From these simulations, we obtained the distribution of the expected mean number of pairs, $f (\bar n_p ; \alpha,\rho)$, for a given angular scale, $\alpha$, and a given density of sources, $\rho$. In the simulated data sets the energy and the arrival direction of the events are randomized according to the experimental uncertainties, and then the mean number of pairs is calculated in the same way as in the real data set.
Then, the value of $\rho_{95}$ which satisfies
\begin{equation}
 \sum_{\bar n_p=0}^{\bar n (\alpha)} f (\bar n_p;  \alpha,\rho_{95})=0.05
\label{eq:bound}
\end{equation}
provides the $95\%$ CL lower bound on the density of sources. Eq. \ref{eq:bound} implies that when comparing the mean number of pairs within a given angle $\alpha$ obtained in a random simulation with source density $\rho_{95}$ to the mean number of pairs of the pseudo data sets, for 95\% of the times the first one will be larger than the second one. Note that the clustering estimator used is the  mean of the number of pairs in the pseudo data sets which is compatible with the experimental uncertainties. This leads to smaller fluctuations in the bound for different realizations of the energy measurement than if just the nominal number of pairs in the events above the threshold is used, as we have checked through numerical simulations.

Bounds will be given for angular scales between $3^\circ$ and $30^\circ$ and for different energy thresholds. Deflections of about $3^\circ$ are likely for 
extragalactic protons of $E> 60$~EeV, and could be larger for strong extragalactic magnetic fields \cite{asm} or for heavier nuclei.
The clustering pattern expected from a particular source scenario may be smoothed out by the deflections introduced by magnetic fields.  
Thus, the bounds obtained at a given angular scale $\alpha$ are only valid if the spread in the arrival directions due to 
 magnetic deflections is smaller than $\alpha$. 

\section{Distribution of sources and simulations
}
\label{subsec:density}

Two plausible scenarios were considered for the spatial distribution of the sources. These were taken to be either uniformly distributed or to follow the local distribution of matter in the universe. 
In both cases, we assumed, for simplicity, equal intrinsic intensity and the same UHECR energy spectrum at the sources. 
The equal intrinsic intensity hypothesis leads to conservative lower bounds on the density of sources, as a dispersion of the intensity of the sources leads typically to a larger number of pairs \cite{dubovsky00}. The presence of magnetic fields along the cosmic ray trajectories, besides leading to deflections in the arrival directions can also produce a magnification or demagnification of the flux received from each particular source. This effect has not been taken into account due to the lack of knowledge of the actual galactic and extragalactic magnetic fields. The magnetic lensing effect is expected to lead to a further dispersion of the apparent intensity of the sources and consequently to a larger number of pairs on average. The reported lower bounds, obtained ignoring this effect, are thus conservative (the interval $\rho > \rho_{95}$ may cover the true value of $\rho$ with a probability larger than 95\%).

\begin{center}
  \begin{figure}[ht]
\epsfig{file=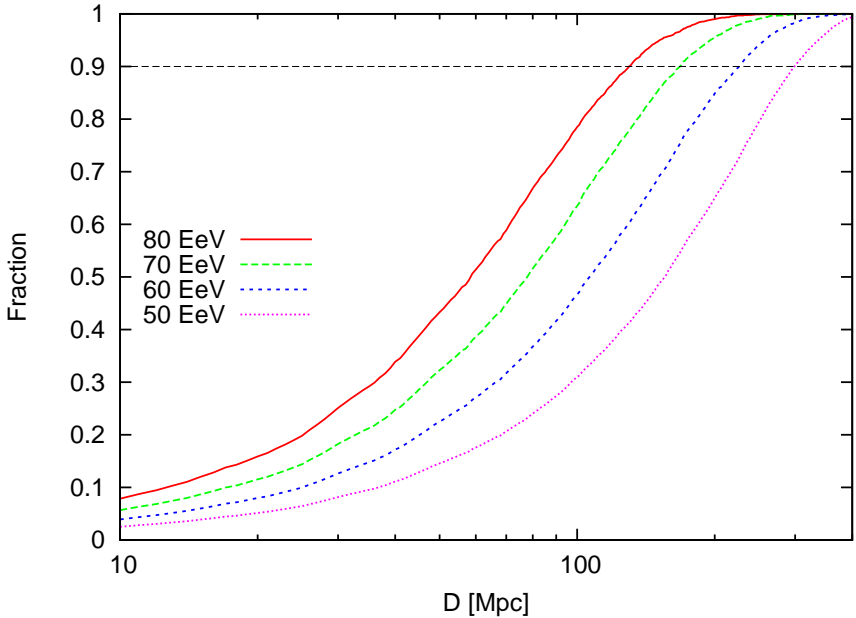,angle=270,width=12cm}
\caption{Fraction of the CR flux coming from distances smaller than D 
for different energy thresholds and a uniform source distribution. 
A spectral index $s=2.2$ was assumed for the proton energy spectrum at emission.
\label{gzk}}
  \end{figure}
\end{center}

Simulations were then performed with the following procedure. For a given density $\rho$, a number $n_s= \rho V$ of sources were homogeneously distributed in a volume $V$. This volume needs to be large enough to originate most of the observed UHECR flux at Earth. An estimate of its size was obtained from Monte Carlo simulations using the CRPropa code \cite{CRPropa}. For a given energy threshold a large number of protons from uniformly distributed sources, with an initial energy spectrum ${\rm d}N/{\rm d}E_i \propto E_i^{-s}$, were followed up to the Earth, taking into account the relevant energy loss processes.  Pion photoproduction  and $e^+e^-$ pair production from the proton interaction with the cosmic microwave background (CMB) were considered. Also, redshift energy losses were included, assuming a $\Lambda$-Cold Dark Matter universe with Hubble constant $H_0=70$ km/s/Mpc, matter energy density $\Omega_m=0.27$ and dark energy density $\Omega_\Lambda=0.73$. The fraction of the flux originated from sources at distances smaller than $D$ is plotted in Figure \ref{gzk} for different energy thresholds for the case of a spectral index $s=2.2$ and a cutoff at $10^{21}$ eV. Its shape was found to change only slightly for values of $s$ between 2 and 2.7  and cutoff energies larger than $10^{20.5}$ eV.

From the simulated sources we propagated protons with initial energy above $E_{\mathrm {min}}=E_{\mathrm {thr}}/1.2$ and a power law spectrum with spectral index 2.2 and recorded all the events arriving to the Earth with $E > E_{\mathrm {min}}$, stopping the simulation after having recorded $N$ events with energy $E > E_{\mathrm {thr}}$. Events were simulated with energies down to $E_{\mathrm {min}}$ so that the energies could be randomized according to the experimental uncertainty as described in the previous section. We simulated sources within radii of 180 Mpc, 230 Mpc and 300 Mpc for $E_{\mathrm {thr}} = 80$ EeV, 70 EeV and 60 EeV respectively, so as to ensure that most of the potential sources were included (see Figure \ref{gzk}). The sources from which the events were propagated were selected with a  probability proportional to $\epsilon(\delta_s,\alpha_s)/D_s^2$, where $\epsilon(\delta_s,\alpha_s)$ is the exposure of the Observatory towards the direction $(\delta_s,\alpha_s)$ of the source and $D_s$ is the distance to the source. In order to account for the energy and angular resolution of the detector, the arrival direction and the energy of the simulated events were randomized according to the corresponding uncertainties. A sample of 2000 sets of $N$ simulated events was used to derive the distribution of the expected mean number of pairs, $f (\bar n_p ; \alpha,\rho)$ (see Section \ref{methods}).

\begin{center}
  \begin{figure}[ht]
\epsfig{file=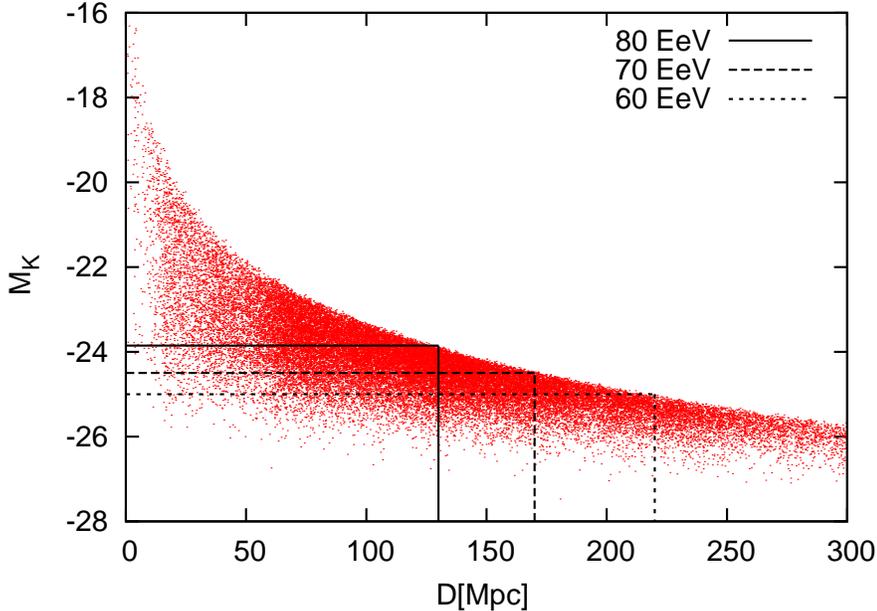,angle=270,width=12cm}
\caption{Absolute magnitude of the galaxies of the 2MRS catalog as a function 
of their distance. The volume limited subsample comprises objects at distances smaller than the vertical lines (solid line, 130 Mpc corresponding to $E_{\mathrm{thr}}=$ 80 EeV; dashed line, 170 Mpc corresponding to $E_{\mathrm{thr}}$= 70 EeV; dotted line, 220 Mpc corresponding to $E_{\mathrm{thr}}$= 60 EeV)  and of magnitude below the corresponding horizontal line. 
\label{cat}}
  \end{figure}
\end{center}
A similar procedure was adopted for the study of sources distributed according to the  2MRS catalog, which maps the 
distribution of matter through near infra-red selected galaxies. 
This catalog 
provides the most densely sampled all-sky survey, covering 91$\%$ of the sky
(excluding a region around the Galactic plane at latitudes $|b| < 5^\circ$ for longitude $|l| > 
30 ^\circ$, and $|b| < 8^\circ$ for $|l| < 30 ^\circ$) \cite{2mass}.
The distance estimated from redshift may be significantly biased by the peculiar velocity for objects with small redshift ($z \lesssim 0.01$). For these objects, we used the distance 
measured independently of the redshift when available in the NASA/IPAC Extragalactic Database (NED).
Otherwise, distances were estimated from the redshift.
The limit in apparent magnitude of the catalog is 
$K =11.75$. To extract volume limited subsamples which 
provide useful tracers of the matter distribution, an absolute magnitude cut 
has to be adopted for any given $D$.
For example, a volume limited subsample for $D = 130$~Mpc (from which $90 \%$ of the flux above 80 EeV is expected to originate) is obtained by requiring the absolute magnitude $M_K$ to be less than  -23.85, as shown in Figure \ref{cat}.
For $D = 170$ Mpc and 220 Mpc (where $90\%$ of the flux above 70 EeV and 60 EeV
is expected to originate),  $M_K < -24.5$ and $M_K < -25$ provide the  volume limited subsamples respectively.
The corresponding densities of objects are 
$\rho_{obj} = 1.5 \times 10^{-3}$Mpc$^{-3}$, $5.9 \times 10^{-4}$Mpc$^{-3}$ and $1.9 \times 10^{-4}$Mpc$^{-3}$
respectively.  Notice that bounds can be reliably placed only up to densities lower than $\rho_{obj}$, since fluctuations will be underestimated in the simulations when sampling from a density of sources  $\rho \simeq \rho_{obj}$. 
We found that bounds for $E_{\mathrm{thr}}= 70$~EeV and 60~EeV 
were too close to the density of objects in the corresponding volume 
limited catalog subsample and thus suffered from the sample variance problem.
For the case of sources distributed like galaxies in the 2MRS catalog, 
we will hence only quote the results for $E_{\mathrm{thr}}= 80$~EeV, where bounds were found to be robust.
In order to  simulate events down to $E_{\mathrm {min}}=E_{\mathrm {thr}}/1.2 \sim 67$ EeV, it is necessary to consider sources up to a distance of $\sim 180$ Mpc. Since the volume limited subsample corresponding to 180 Mpc would lead to a too  low density of objects we use the cut in magnitude corresponding to 130 Mpc ($M_k<-23.85$) but include all the objects in the catalog up to 180 Mpc. We compensate the relative lack of galaxies at distances larger than 130 Mpc by including a weighting factor $F(D_S)$ for the sources with $D_s > 130$ Mpc. This weighting factor is inversely proportional to the selection function of the catalog
\footnote{This is the probability of detecting a galaxy in the survey as a function of the distance, $\Phi(r) \propto r^{-2} dn_s/dr$, with $dn_s/dr$ the distribution of objects as a function of the distance in the catalog.}.
To determine $f (\bar n_p ; \alpha,\rho)$, Monte Carlo simulations were performed following a procedure analogous to that of the isotropic distribution of sources, but with sources drawn from the subsample of the catalog. The only difference is that the sources are selected from the catalog subsample with a probability proportional to $F(D_s) \epsilon(\delta_s,\alpha_s)/D_s^2$, with $F(D_s)=1$ for $D_s<$ 130 Mpc and $F(D_s)=\Phi(\mathrm {130 Mpc})/\Phi(D_s)$ for $D_s >$ 130 Mpc.

Notice that the propagation of particles was performed under the assumption that UHECRs are protons. As a matter of fact, the propagation for iron nuclei and its secondaries, mainly determined by nuclear photodisintegration through interaction with the CMB and infrared background, leads to a very similar attenuation of the flux as a function of the energy.  Intermediate mass primary nuclei experience larger energy losses \cite{hmr06}, 
 and can thus reach the Earth only if produced quite nearby. As a consequence, the expected  clustering is higher than that of the proton or iron cases, 
and the 95\% CL bounds for intermediate mass nuclei UHECRs are hence expected to be tighter than those obtained in this paper assuming a pure proton composition.

\section{Results}\label{results}

In this section we derive bounds on the density of sources in the nearby 
universe, following the procedures detailed in the previous sections and focusing first in the case of uniformly distributed sources. We present the detailed analysis for a threshold of 70 EeV and then show the results for the 60 EeV and 80 EeV thresholds.
In the two-point correlation analysis, a higher threshold energy reduces the maximum distance to Earth traveled by UHECRs, leading to a stronger discrimination of the clustering signal. On the other hand, the number of selected events may be drastically reduced, introducing large statistical uncertainties in the derived bounds. Based on numerical simulations we found that the 70 EeV threshold represents the preferred balance for the present statistics.
 
\begin{figure}[ht]
\epsfig{file=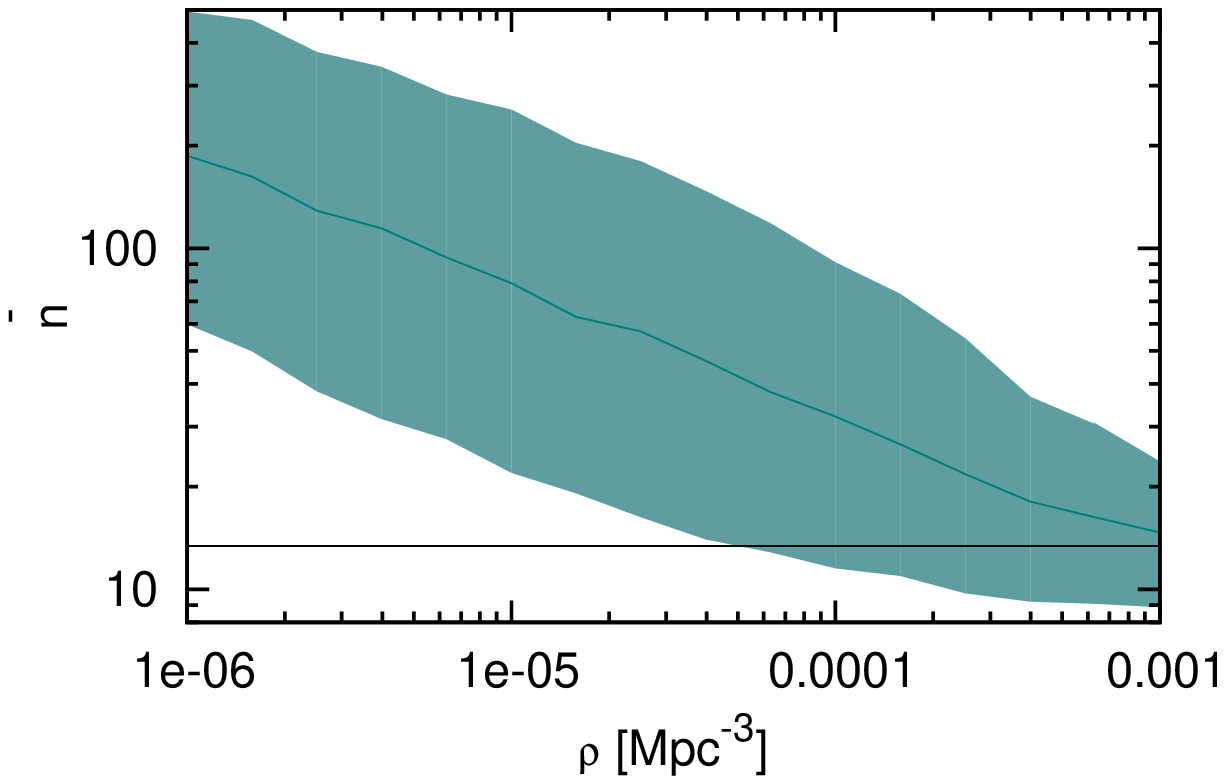,angle=270,width=7cm}
\epsfig{file=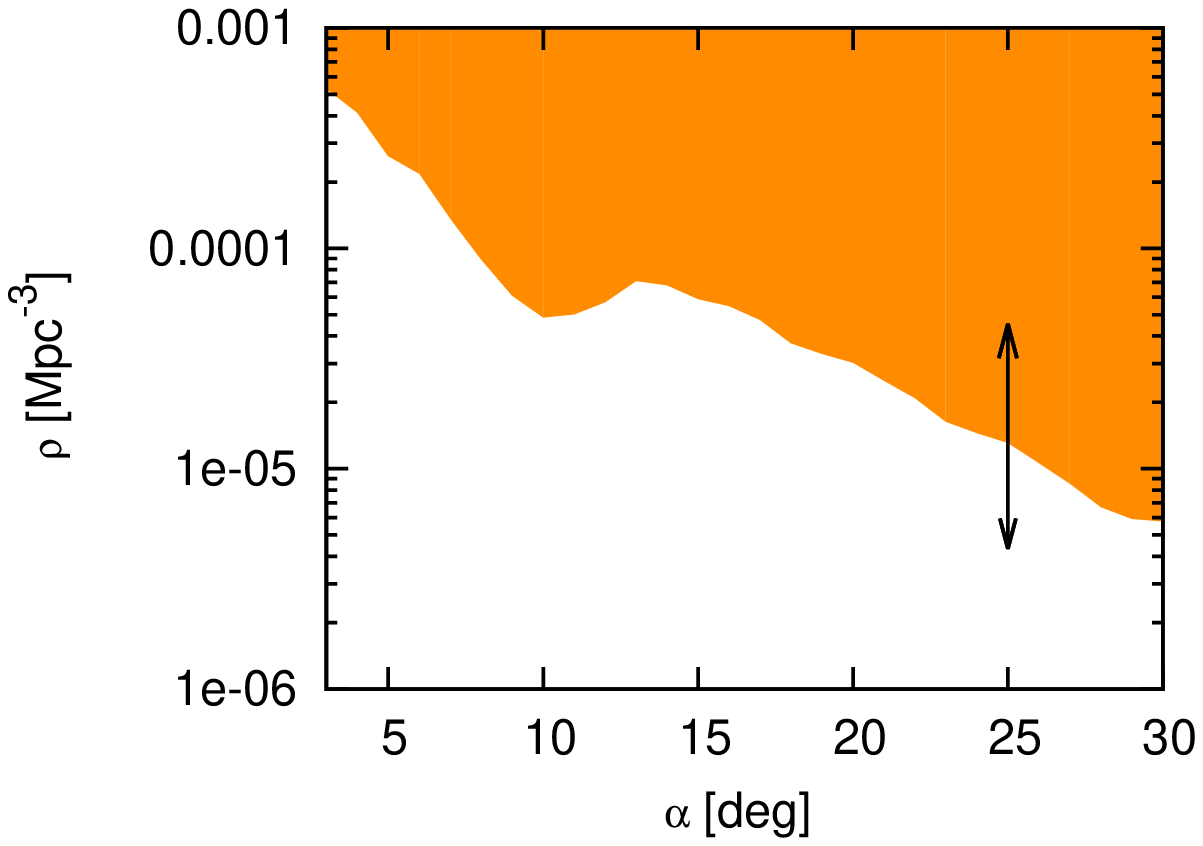,angle=270,width=7cm}
\caption{Results for uniformly distributed sources derived  for $E_{\mathrm{thr}}=70$~EeV (43 highest energy events). Left: The mean number of pairs at an angular scale of $10^\circ$.  The blue line and the shaded band represent the mean and $90\%$ CL limits on the expected number of pairs from Monte Carlo simulations. The mean number of pairs for the data is indicated by the solid horizontal line.
Right: $95\%$ CL allowed region (shaded area) for the density of sources as a function of the 
angular scale. The vertical arrows indicate how much the bounds change for 
a 22\% shift of the absolute energy scale. 
\label{70EeV}}
\end{figure}

In Fig. \ref{70EeV} we present the results for $E_{\mathrm{thr}}=70$~EeV, corresponding to 43 selected events.
To illustrate the method, the mean number of pairs at an angular scale 
$\alpha=10^\circ$, ${\bar n}(10^\circ )$, is shown as a function of the density of sources (left panel). For any given density $\rho$,
the distribution of the expected mean number of pairs, $f ({\bar n}_p ; 10^\circ,\rho)$, is obtained from simulations (see Section \ref{subsec:density}). The shaded band in the left panel of Fig.  \ref{70EeV} represents the 90\% CL of $f ({\bar n}_p ; 10^\circ,\rho)$, with $5\%$ of the time the mean number of pairs being above the band and $5\%$ of the time below it. The value of the mean number of pairs obtained for the data is indicated by the solid horizontal line. A 95\% CL lower bound on the density of sources at  $\alpha=10^\circ$ is then obtained from Eq. \ref{eq:bound}, corresponding to the value of the density for which the lower end of the band and the horizontal line intersect.
Bounds at other angular scales are derived with an analogous procedure.
We emphasize again that bounds at a given angular scale $\alpha$ are only valid if the spread in the cosmic ray arrival directions due to magnetic deflections is smaller than $\alpha$. Thus, we present in the right panel of Fig.  \ref{70EeV} results for angular scales between $3^\circ$ and $30^\circ$, which cover a wide range of potential deflections due to magnetic fields and for different UHECR composition.

The most stringent bound is obtained for $\alpha=3^\circ$, where the density of sources is found to be larger than $5 \times 10^{-4}$ Mpc$^{-3}$ with $95\%$ CL. For larger angular scales, the bound is less restrictive, reaching $6 \times 10^{-6}$ Mpc$^{-3}$ for $\alpha=30^\circ$.
 The vertical arrows indicate how much the bounds change when the absolute energy scale of the experiment is shifted by $\pm 22\%$ according to its systematic uncertainty. 
 To estimate this effect, we assumed that the true energy threshold for the 43 selected events was $E_{\mathrm{thr}}= 55$ or 85~EeV, rather than the nominal 70~EeV, and repeated the procedure to set the lower bounds. We found that a $22\%$ upward (downward) shift in energy moves the bounds upward (downward) by about a factor of 3, as indicated by the vertical arrows.
\begin{figure}[ht]
\epsfig{file=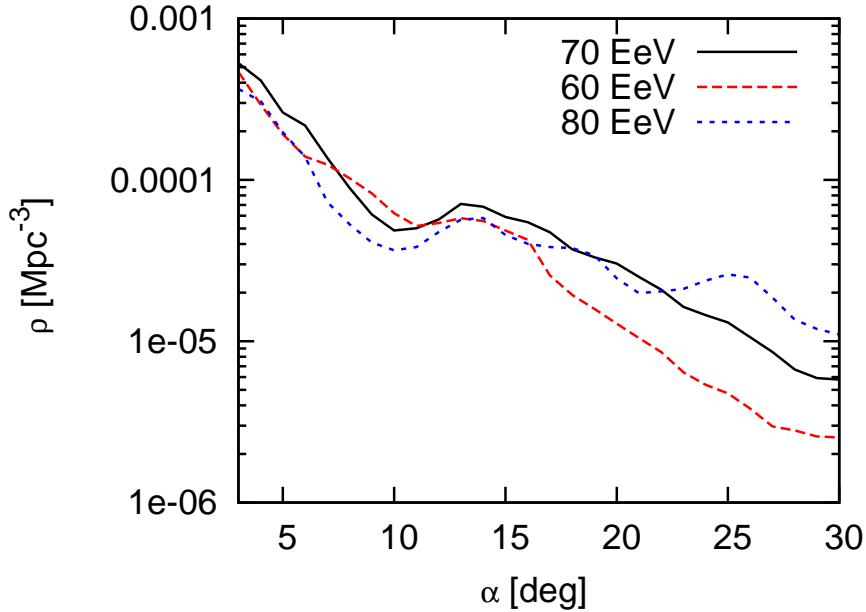,angle=270,width=12cm}
\caption{$95\%$ CL lower bounds on the density of uniformly distributed sources as a function of the angular scale. Bounds derived with $E_{\mathrm{thr}}= 60,~70$ and 80~EeV are shown.
\label{60-70-80EeV}}
\end{figure}

The $95\%$ CL lower bounds on the density of an isotropic distribution of sources for $E_{\mathrm{thr}}= 60$~EeV ($N=84$ events) and  80~EeV ($N=22$ events)
are given in Figure \ref{60-70-80EeV}, together with the bounds at $E_{\mathrm{thr}}= 70$~EeV. It can be seen that the bounds are quite stable with respect to the energy threshold choice.

If the intrinsic intensity of the sources were not uniform, a larger clustering of events is typically expected and thus tighter bounds on the density of sources would result. We have checked that for a distribution of intensities with dispersion equal to the mean the bound is shifted up by $\sim 50\%$.
\begin{figure}[ht]
\epsfig{file=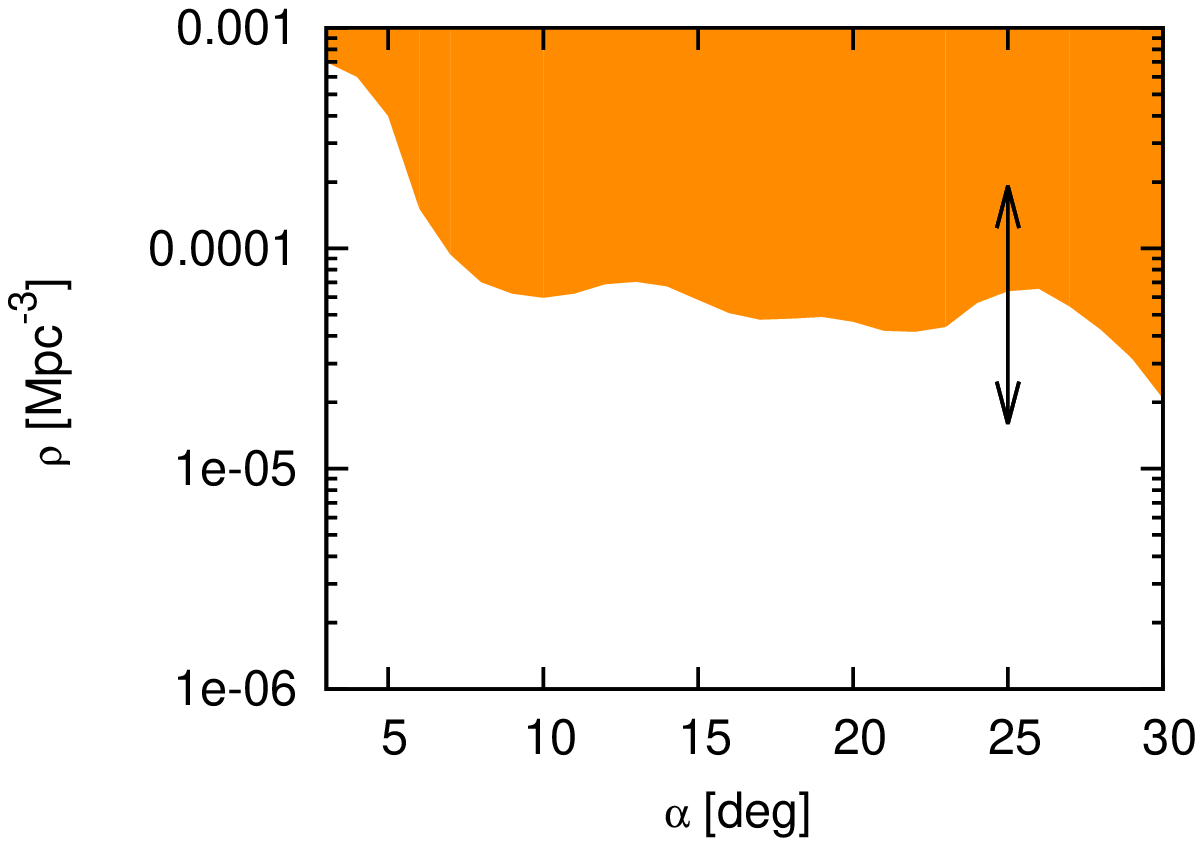,angle=270,width=12cm}
\caption{Bounds on a distribution of sources following the local matter density traced by the 2MRS catalog galaxies, derived with 20 events above $E_{\mathrm{thr}}=80$~EeV. The shaded area represents the $95\%$ CL allowed region for the density of sources. The vertical arrows indicate the uncertainty associated to a $22 \%$ uncertainty in the energy calibration.
\label{2mrs80EeV}}
\end{figure}

Stronger bounds on the density of sources are expected to result when the sources are not uniformly distributed in space, due to the 
additional clustering of the sources themselves.
We explored the possibility that UHECR sources follow the distribution of matter in the local universe by using the galaxies in the 2MRS catalog as tracers of the matter distribution. Since the Galactic plane is masked in the 2MRS catalog, cosmic rays coming from that region of the sky are not included in this study,  which results in $N=20$ events above an energy threshold of 80~EeV. This procedure ensures that simulated and real events can arrive from the same solid angle of the sky. Due to deflections in the regular magnetic field the sources of events in the unmasked region may lie in the mask. We assume that the clustering properties of the sources in both regions are similar.
To derive bounds on the density of sources, we followed the same procedure described for the case of uniformly distributed sources, but with sources drawn from the volume limited subsample of the catalog (see Section \ref{subsec:density}). The corresponding $95\%$ CL lower bound as a function of the angular scale is shown in Figure \ref{2mrs80EeV}, ranging from 
$7 \times 10^{-4}$ Mpc$^{-3}$ at $\alpha=3^\circ$ to $2 \times 10^{-5}$ Mpc$^{-3}$ at $\alpha=30^\circ$.
A $\pm 22\%$ shift of the absolute energy scale shifts the bounds as indicated by the vertical arrows.
  
In a previous study  \cite{agnauger10}, the arrival directions of events above 55 EeV detected by the Pierre Auger Observatory were found to be compatible with a model based on 2MRS galaxies for smoothing angles of a few degrees and correlating fractions of about 40\%. These parameters are however not strongly constrained with the present statistics. The clustering analysis considered here represents a complementary study that probes the density of the sources.

\section{Conclusions}\label{conclusions}

We have used the two-point angular correlation function to study the clustering properties of the arrival directions of UHECRs detected by the Pierre Auger Observatory.
No significant excess of pairs up to an angular scale of $30^\circ$ was found, which provides a lower bound on the density of UHECR sources.  

From the analysis of events with energy above 70 EeV we found that, if the spread due to magnetic deflections is smaller than $\alpha = 3^\circ$, the density of  equal intrinsic intensity sources uniformly distributed in space is larger than $5.3 \times 10^{-4}$ Mpc$^{-3}$ with $95\%$ CL. For larger angular scales, the bound is less restrictive, reaching $6 \times 10^{-6}$ Mpc$^{-3}$ at $\alpha=30^\circ$. These bounds have a factor of 3 uncertainty arising from the 22\% systematic uncertainty in the energy scale.
The analysis of events with energy larger than 60 and 80~EeV yielded comparable limits. 
We also studied a distribution of UHECR sources following the local matter in the universe, which was traced by the 2MRS catalog of galaxies. Bounds on the density of sources were similar, ranging from $7 \times 10^{-4}$ Mpc$^{-3}$ at $\alpha = 3^\circ$ up 
to $2 \times 10^{-5}$ Mpc$^{-3}$ at $\alpha = 30^\circ$.
Since the spread in the UHECR arrival directions induced by magnetic fields may wash out the clustering signal, a physical interpretation of these bounds is meaningful only when the spread due to magnetic deflections is smaller than the angular scale. 
 
Even with this limitation, our bounds provide novel insight into the origin of UHECRs.  
If magnetic deflections are limited to a few degrees, as expected for a light composition of cosmic rays at the highest energies and weak extragalactic magnetic fields \cite{dolag}, our data suggest a rather large value for the source density, $\rho> 10^{-4}$ Mpc$^{-3}$. This value is for instance compatible with the density of galaxies brighter than $10^{11} L_\odot$ \cite{soifer86}, 
but disfavors bright radio galaxies and AGNs with $L_X > 10^{43}$ erg $s^{-1}$ as the main source of the observed flux of cosmic rays above 70~EeV. 
On the other hand, lower values of the density are allowed for large magnetic deflections, as expected for a predominantly heavy composition or  stronger extragalactic magnetic fields. In this case, the observed clustering is still compatible with the density of some types of AGNs,
like Seyfert galaxies or low luminosity, Fanaroff-Riley I, radio galaxies.

\bigskip\bigskip

This paper has made use of the NASA/IPAC Extragalactic Database (NED) 
which is operated by the Jet Propulsion Laboratory, California Institute of 
Technology, under contract with the National Aeronautics and Space 
Administration. 

The successful installation, commissioning, and operation of the Pierre Auger Observatory
would not have been possible without the strong commitment and effort
from the technical and administrative staff in Malarg\"ue.

We are very grateful to the following agencies and organizations for financial support: 
Comisi\'on Nacional de Energ\'ia At\'omica, 
Fundaci\'on Antorchas,
Gobierno De La Provincia de Mendoza, 
Municipalidad de Malarg\"ue,
NDM Holdings and Valle Las Le\~nas, in gratitude for their continuing
cooperation over land access, Argentina; 
the Australian Research Council;
Conselho Nacional de Desenvolvimento Cient\'ifico e Tecnol\'ogico (CNPq),
Financiadora de Estudos e Projetos (FINEP),
Funda\c{c}\~ao de Amparo \`a Pesquisa do Estado de Rio de Janeiro (FAPERJ),
Funda\c{c}\~ao de Amparo \`a Pesquisa do Estado de S\~ao Paulo (FAPESP),
Minist\'erio de Ci\^{e}ncia e Tecnologia (MCT), Brazil;
AVCR AV0Z10100502 and AV0Z10100522, GAAV KJB100100904, MSMT-CR LA08016,
LG11044, MEB111003, MSM0021620859, LA08015, TACR TA01010517 and GA UK 119810, Czech Republic;
Centre de Calcul IN2P3/CNRS, 
Centre National de la Recherche Scientifique (CNRS),
Conseil R\'egional Ile-de-France,
D\'epartement  Physique Nucl\'eaire et Corpusculaire (PNC-IN2P3/CNRS),
D\'epartement Sciences de l'Univers (SDU-INSU/CNRS), France;
Bundesministerium f\"ur Bildung und Forschung (BMBF),
Deutsche Forschungsgemeinschaft (DFG),
Finanzministerium Baden-W\"urttemberg,
Helmholtz-Gemeinschaft Deutscher Forschungszentren (HGF),
Ministerium f\"ur Wissenschaft und Forschung, Nordrhein-Westfalen,
Ministerium f\"ur Wissenschaft, Forschung und Kunst, Baden-W\"urttemberg, Germany; 
Istituto Nazionale di Fisica Nucleare (INFN),
Ministero dell'Istruzione, dell'Universit\`a e della Ricerca (MIUR), Italy;
Consejo Nacional de Ciencia y Tecnolog\'ia (CONACYT), Mexico;
Ministerie van Onderwijs, Cultuur en Wetenschap,
Nederlandse Organisatie voor Wetenschappelijk Onderzoek (NWO),
Stichting voor Fundamenteel Onderzoek der Materie (FOM), Netherlands;
Ministry of Science and Higher Education,
Grant Nos. N N202 200239 and N N202 207238, Poland;
Portuguese national funds and FEDER funds within COMPETE - Programa Operacional Factores de Competitividade through 
Funda\c{c}\~ao para a Ci\^{e}ncia e a Tecnologia, Portugal;
Romanian Authority for Scientific Research ANCS, 
CNDI-UEFISCDI partnership projects nr.20/2012 and nr.194/2012, 
project nr.1/ASPERA2/2012 ERA-NET and PN-II-RU-PD-2011-3-0145-17, Romania; 
Ministry for Higher Education, Science, and Technology,
Slovenian Research Agency, Slovenia;
Comunidad de Madrid, 
FEDER funds, 
Ministerio de Ciencia e Innovaci\'on and Consolider-Ingenio 2010 (CPAN),
Xunta de Galicia, Spain;
Science and Technology Facilities Council, United Kingdom;
Department of Energy, Contract Nos. DE-AC02-07CH11359, DE-FR02-04ER41300, DE-FG02-99ER41107,
National Science Foundation, Grant No. 0450696,
The Grainger Foundation USA; 
NAFOSTED, Vietnam;
Marie Curie-IRSES/EPLANET, European Particle Physics Latin American Network, 
European Union 7th Framework Program, Grant No. PIRSES-2009-GA-246806; 
and UNESCO.


\begin{thebibliography}{99}

 \bibitem{GZK} K.~Greisen, Phys.\ Rev.\ Lett.\  16 (1966) 748;
G.\ T.\ Zatsepin and V.\ A.\ Kuz'min, JETP Lett. 4 (1966) 78. 

 \bibitem{spechires} R. U. Abbasi et al. (HiRes Collaboration), 
Phys. Rev. Lett. 100 (2008) 101101.

 \bibitem{specauger} J.~Abraham et al. (The Pierre Auger Collaboration), 
Phys. Rev. Lett. 100 (2008) 211101; 
J.~Abraham et al. (The Pierre Auger Collaboration),
 Phys. Lett. B 685 (2010) 239.

\bibitem{kochanek01} C.~S.~Kochanek et al., Astrophys. J. 560 (2001) 566.

\bibitem{agndens} A.~T.~Steffen et al., Astrophys. J. Lett. 596 (2003) L3;
S.~Sazonov et al., Astron. and Astrophys. J. 462 (2007) 57; J.~Tueller et al.,  
Astrophys. J. 681 (2008) 113.

\bibitem{carlberg97} R.~G.~Carlberg et al., Astrophys. J. Lett. 479 (1997) L19.

 \bibitem{dolag} K.~Dolag et al. JCAP 0501 (2005) 009.

 \bibitem{asm} E.~Armengaud, G. Sigl and F. Miniati, Phys. Rev. D 72 (2005)
043009. 

\bibitem{augerxmax} J.~Abraham et al. (The Pierre Auger Collaboration), 
Phys. Rev. Lett. 104 (2010) 091101.

 \bibitem{augercomp} P.~Facal San Luis for the Pierre Auger Collaboration,
Proc. 32nd ICRC (2011) [arXiv:1107.4804].

\bibitem{icrccomp} D.~Garcia-Pinto for the Pierre Auger Collaboration,
Proc. 32nd ICRC (2011) [arXiv:1107.4804].

 \bibitem{hirescomp}  R. U. Abbasi et al. (HiRes Collaboration), 
Phys. Rev. Lett. 104 (2010) 161101.

 \bibitem{tacomp} C. Jui et al. (TA Collaboration) Proc. APS DPF Meeting
[arXiv:1110.0133].

 \bibitem{dubovsky00}
S. L. Dubovsky, P. G. Tinyakov and I. I. Tkachev, Phys. Rev. Lett. 85 (2000)
1154.

 \bibitem{fodor00}
Z. Fodor and S. D. Katz, Phys. Rev. D 63 (2000) 23002.

 \bibitem{yoshi03}
H. Yoshiguchi et al., Astrophys. J. 586 (2003) 1211.

 \bibitem{blasi04}
P. Blasi and S. De Marco, Astrop. Phys. 20 (2004) 559.

\bibitem{sigl04}
G.~Sigl, F.~Miniati and T.~A.~En{\ss}lin, Phys. Rev. D 70 (2004) 043007.

 \bibitem{kachel05}
M. Kachelriess and D. Semikoz, Astrop. Phys. 23 (2005) 486.

 \bibitem{cuoco09}
A. Cuoco et al., Astrophys. J.  702 (2009) 825.

\bibitem{takami}
H.~Takami and K.~Sato, Astrop. Phys. 30 (2009) 306.

\bibitem{ta2012}
T. Abu-Zayyad et al. (TA Collaboration), Astrophys. J. 
757 (2012) 26.

\bibitem{2mass}
J.~P.~Huchra et al., Astrophys. J. Suppl. 199 (2012) 26.

\bibitem{dedomenico}
M.~de Domenico for the Pierre Auger Collaboration,
Proc. 32nd ICRC (2011) [arXiv:1107.4805].

\bibitem{augernim}
J.~Abraham et al. (The Pierre Auger Collaboration), NIM  A 523 (2004) 50.

\bibitem{augersup}
I.~Allekotte  et al. (The Pierre Auger Collaboration), NIM A 586 (2008) 409.

\bibitem{augerfl}
J.~Abraham et al. (The Pierre Auger Collaboration), NIM A 620 (2010) 227.

\bibitem{angres2009}
C.~Bonifazi, for The Pierre Auger Collaboration, Nucl. Phys. B (Proc. Suppl.)
190 (2009) 20.

\bibitem{enres2011}
R. Pesce, for The Pierre Auger Collaboration, Proc. 32nd ICRC (2011) [arXiv:1107.4809].

\bibitem{CRPropa} 
K.-H.~Kampert et al., [arXiv:1206.3132].

\bibitem{cowan}
G.~Cowan, Review of Particle Physics, Phys. Rev. D 86, 010001 (2012) 390.

\bibitem{hmr06} 
D.~Harari, S.~Mollerach and E.~Roulet, JCAP 11(2006)012.

\bibitem{agnauger10}
P.~Abreu et al.  (Pierre Auger Collaboration), Astropart. Phys. 34 (2010) 314.

\bibitem{soifer86} B.~T.~Soifer et al., Astrophys. J. Lett. 303 (1986) L41.

\end{thebibliography}
\end{document}